\documentclass{article}
\usepackage[final]{graphicx}

\begin{document}

\title{Non-unitary Process and Quantum Communication}
\author{Riuji Mochizuki\thanks{E-mail:rjmochi@tdc.ac.jp}\\
Laboratory of Physics, Tokyo Dental College,\\ 2-9-7 Kandasurugadai, Chiyoda-ku, Tokyo 101-0062, Japan }
\maketitle
\begin{abstract}%
The no-communication theorem states that the observation of a subsystem of an entangled state does not affect another subsystem. Nevertheless, this theorem is based on the assumption that all quantum processes are unitary. We examine a feasible thought experiment and show that a non-unitary process included in this thought experiment enables the transmission of information by means of the quantum observation process.

\end{abstract}

The development of quantum states obeying the Schr{\" o}dinger equation, which is one of the fundamental equations in quantum mechanics, is unitary.  In the Copenhagen interpretation, it is assumed that the quantum mechanics includes a non-unitary process, which is the apparent collapse of the wave function. Although it is of considerable practical use, there remain many theoretical difficulties. To overcome these, a number of quantum observation theories without non-unitary processes ({\it unitary quantum mechanics}) were suggested\cite{eve, DeW, Zeh2, Sch, Sau, Zeh1, Kub, Zurek1}. However, imperfections in unitary quantum mechanics have been observed\cite{Healey}. Moreover, it has been shown that non-unitarity is introduced naturally into quantum measurement processes if the uncertainty relation is taken into account\cite{mochi1}.

The question arose whether there were experiments that could determine if quantum mechanics was unitary. For example, we examine the EPR-Bohm experiment\cite{EPR, Bohm}. Suppose there are a pair of observers {\it Anne} and {\it Bill}. First, Anne measures the spin of particle 1 in direction {\it a}; then, Bill measures the spin of particle 2 in direction {\it b}. We define unitary operators $\hat A_a$ and $\hat B_b$ corresponding to the measurement processes of Anne and Bill, respectively. The density matrix $\hat \rho_{ab}$ of the system after their operation is written as
\begin{equation}
\hat\rho_{ab}=\hat B_b\hat A_a|I\rangle\langle I|\hat A^{\dag}_a\hat B^{\dag}_b,\label{eq:appb2}
\end{equation}
where $|I\rangle$ is the initial state of the system.  Due to the unitarity
\begin{equation}
\hat A^{\dag}_a\hat A_a={\bf 1}\label{eq:unitarity2}
\end{equation}
and commutativity
\[
[\hat A_a,\hat B_b]=0,
\]
we can trace out $\hat A_a$ from the density matrix (\ref{eq:appb2}) to obtain the reduced density matrix $\hat\rho_b$, which represents the probability distribution of Bill's measurement 
\begin{equation}
\hat\rho_b=\hat B_b|I\rangle\langle I|\hat B^{\dag}_b.\label{eq:nc1}
\end{equation}
As (\ref{eq:nc1}) includes no information about Anne's operation, Bill would not know what Anne did.  Therefore, it is impossible for Anne to provide any information by means of her measurement process {\it if her operation does not result in anything non-unitary}. 

Taking the contraposition of the no-communication theorem, we can conclude that quantum mechanics includes some non-unitarity if communication via the quantum observation process (CQOP) is possible. However, there is a current of opinion is that the special relativity theory prohibits such communication.  As shown by a number of experiments\cite{Aspect1,Aspect2,Aspect3,GM,MWZ, Nat}, the quantum world violates the Bell inequality\cite{Bell} and other Bell-type inequalities\cite{CH1,FC,CH2}. This indicates that quantum mechanics cannot be described by a local hidden variable theory.  Nevertheless, the non-locality of quantum mechanics does not ensure that CQOP is instantaneous, if it exists.  Recently, it was shown by means of the von Neumann-type measurement model\cite{von} that slow (i.e. at the speed of light or slower) CQOP is possible and consistent with both the theory of special relativity and non-locality of quantum mechanics\cite{mochi2}.

We now examine a feasible thought experiment; independently of any observation models, we show that slow CQOP is possible. We consider an experiment wherein the spin of an electron $S$ is observed. $|+\rangle$ and $|-\rangle$ are the eigenstates of its spin in the $z$ direction belonging to its eigenvalues $+\hbar/2$ and $-\hbar/2$, respectively. 

The initial state $|r\rangle$ of $S$ is the eigenstate of $\hat\sigma_x$
\begin{equation}
\hat\sigma_x=|+\rangle\langle-|+|-\rangle\langle +|
\end{equation}
belonging to its eigenvalue 1:
\begin{equation}
|r\rangle=\frac{1}{\sqrt{2}}\Big(|+\rangle+|-\rangle\Big),\label{eq:appb3}
\end{equation}
\[
\hat\sigma_x|r\rangle=|r\rangle.
\]
The initial state of the probe $M_A$ of an observer ({\it Alice}) is $|a_0\rangle$, which changes due to its interaction with $S$ as
\begin{equation}
|+\rangle\otimes |a_0\rangle\ \rightarrow\ |+\rangle\otimes |a_+\rangle,
\end{equation}
\begin{equation}
|-\rangle\otimes |a_0\rangle\ \rightarrow\ |-\rangle\otimes |a_-\rangle,
\end{equation}
where $\langle a_+|a_-\rangle =0$.  
In contrast, the initial state of the probe $M_B$ of another observer ({\it Bob}) is $|b_0\rangle$, which changes due to its interaction with $S$ as
\begin{equation}
|r\rangle\otimes |b_0\rangle\ \rightarrow\ |r\rangle\otimes |b_r\rangle,
\end{equation}
\begin{equation}
|l\rangle\otimes |b_0\rangle\ \rightarrow\ |l\rangle\otimes |b_l\rangle,
\end{equation}
where $\langle b_r|b_l\rangle =0$ and
\[
|l\rangle=\frac{1}{\sqrt{2}}\Big(|+\rangle -|-\rangle\Big).
\]
The initial state of the unified system is
\begin{equation}
|\Phi\rangle\equiv |r\rangle\otimes|a_0\rangle\otimes|b_0\rangle.\label{eq:shoki}
\end{equation}
%???

First, $M_A$ interacts with $S$ and the state of the unified system becomes
\begin{equation}
\begin{array}{rl}
|\Phi\rangle \rightarrow |\Phi_0\rangle &= \hat O |\Phi\rangle \\
&=\frac{1}{\sqrt{2}}\Big(|+\rangle\otimes |a_+\rangle+|-\rangle\otimes |a_-\rangle\Big)\otimes |b_0\rangle\\
&=\frac{1}{2}\Big[\big(|r\rangle+|l\rangle\big)\otimes|a_+\rangle+\big(|r\rangle-|l\rangle\big)\otimes|a_-\rangle\Big]\otimes|b_0\rangle,\label{eq:Alice1}
\end{array}
\end{equation}
where
\begin{equation}
\hat O\equiv \Big(|+\rangle\langle +|\otimes|a_+\rangle\langle a_0|+|-\rangle\langle -|\otimes|a_-\rangle\langle a_0|\Big)\otimes {\bf 1},
\end{equation}

Next, Alice carries out either of the following two operations (I or II) {\it only} on $M_A$ to transmit information to Bob. Subsequently, Bob operates $M_B$ to interact with $S$ to receive the information.\\
I. Alice returns the state of $M_A$ to $|a_0\rangle$ to erase the information of the spin of $S$ from $M_A$, and the unified system returns to its initial state (\ref{eq:shoki}):
\begin{equation}
\begin{array}{rl}
|\Phi_0\rangle\rightarrow|\Phi_I\rangle&=\hat A\hat O|\Phi\rangle\\
&=|\Phi\rangle,\label{eq:AI}
\end{array}
\end{equation}
where
\begin{equation}
\hat A\equiv {\bf 1}\otimes\Big(|a_0\rangle\langle a_+|+|a_0\rangle\langle a_-|\Big)\otimes{\bf 1},\label{eq:AI}
\end{equation}
Then, $M_B$ interacts with $S$:
\begin{equation}
\begin{array}{rl}
|\Phi_I\rangle\rightarrow|\Phi^{\prime}_I\rangle&=\hat B|\Phi_I\rangle\\
&=|r\rangle\otimes|a_0\rangle\otimes|b_r\rangle,\label{eq:phiprime}
\end{array}
\end{equation}
where
\[
\hat B\equiv|r\rangle\langle r|\otimes{\bf 1}\otimes |b_r\rangle\langle b_0|+|l\rangle\langle l|\otimes{\bf 1}\otimes |b_l\rangle\langle b_0|.
\]
The expectation value of the output of Bob's measuring device for ({\ref{eq:phiprime}) is
\begin{equation}
\langle \Phi^{\prime}_I|\hat\beta|\Phi^{\prime}_I\rangle =1,\label{eq:Ib}
\end{equation}
where
\begin{equation}
\hat\beta\equiv{\bf 1}\otimes{\bf 1}\otimes\Big(|b_r\rangle\langle b_r|-|b_l\rangle\langle b_l|\Big).
\end{equation}
\\
II. Alice does nothing so that $M_A$ retains the information of the spin of $S$.  Then, $M_B$ interacts with $S$:
\begin{equation}
\begin{array}{rl}
|\Phi_0\rangle\rightarrow |\Phi^{\prime}_0\rangle&=\hat B|\Phi_0\rangle\\
&=\frac{1}{2}\Big(|r\rangle\otimes|a_+\rangle\otimes|b_r\rangle+|l\rangle\otimes|a_+\rangle\otimes|b_l\rangle\\
&\ \ +|r\rangle\otimes|a_-\rangle\otimes|b_r\rangle-|l\rangle\otimes|a_-\rangle\otimes|b_l\rangle\Big).\label{eq:phi0prime}
\end{array}
\end{equation}
In contrast to (\ref{eq:Ib}), the expectation value of the output of Bob's measuring device for ({\ref{eq:phi0prime}) is
\begin{equation}
\langle\Phi^{\prime}_0|\hat\beta|\Phi^{\prime}_0\rangle=0.\label{eq:IIb}
\end{equation}

Thus, we conclude that Alice can change the expectation value of the spin measured by Bob by means of the operation only on her device. The reason we have arrived at this conclusion, despite the no-communication theorem, is that the operation carried out by Alice in I is not unitary, i.e. the operator (\ref{eq:AI}) is not unitary but satisfies
\[
\hat A\Big[|S\rangle\otimes\big(|a_+\rangle -|a_-\rangle\big)\otimes |B\rangle\Big]=0,
\]
where $|S\rangle$ and $|B\rangle$ are arbitrary states of $S$ and $M_B$, respectively.
In other words, if this operation were to be unitary, information regarding the spin of $S$ would not be lost from Alice's device. We are convinced that such an operation is feasible owing to the experiments on the quantum eraser\cite{eraser1, Kim}. Moreover, this series of operations does not conflict with the no-cloning theorem\cite{Woo, Dieks} or no-deleting theorem\cite{Pati}, which forbid making separable\cite{Esp} copies of a state or deleting separable states; however, they do not restrict the entangling of states or undoing of entangled states.

In summary, we demonstrated that the transmission of information by means of a quantum observation process is possible. Conversely, taking the contraposition of the no-communication theorem, we concluded that quantum mechanics needs some non-unitary processes if such communication is shown to be possible by experiments.

\end{document}